\documentclass[journal]{IEEEtran}
\usepackage{color}
\pdfoutput=1 
\ifCLASSINFOpdf
   \usepackage[pdftex]{graphicx}
   \DeclareGraphicsExtensions{.pdf,.jpeg,.png}
\else
\fi
\usepackage{array}

\ifCLASSOPTIONcompsoc
  \usepackage[caption=false,font=normalsize,labelfont=sf,textfont=sf]{subfig}
\else
  \usepackage[caption=false,font=footnotesize]{subfig}
\fi
\usepackage{url}
\usepackage{makecell}

\usepackage{multirow}
\usepackage{mathtools}
\usepackage{bm}

\usepackage[tableposition=top]{caption}
\usepackage{listings}
\newcommand\blfootnote[1]{%
  \begingroup
  \renewcommand\thefootnote{}\footnote{#1}%
  \addtocounter{footnote}{-1}%
  \endgroup
}

\begin{document}
\title{Omni SCADA Intrusion Detection Using Deep Learning Algorithms}

\author{Jun~Gao, Luyun~Gan, Fabiola~Buschendorf, Liao~Zhang, Hua~Liu, Peixue~Li, Xiaodai~Dong and Tao~Lu}
\markboth{}{Omni SCADA Intrusion Detection Using Deep Learning Algorithms}
\maketitle
\blfootnote{Manuscript received June 14, 2019; revised XY, 2019.This work is supported in part by the Nature Science and Engineering Research Council of Canada (NSERC) Discovery Grant (Grant No. RGPIN-2015-06515), Mitacs globalink program, and Nvidia
Corporation TITAN-X GPU grant. (Corresponding author: Tao Lu)}
\blfootnote{J.~Gao, L.~Gan, L.~Zhang, X.~Dong and T.~Lu  are with the Department of Electrical and Computer Engineering, University of Victoria, EOW 448, 3800 Finnerty Rd., Victoria, British Columbia, V8P 5C2, Canada, (e-mail: \{jungao,luyun,liao,xdong,taolu\}@uvic.ca)}
\blfootnote{F.~Buschendorf was with the Department
of Computer Science, University of Goettingen, Germany, (e-mail: fabiola.buschendorf@protonmail.com)}
\blfootnote{H.~Liu and P.~Li are with Fortinet Technology Inc., 899 Kifer Road, Sunnyvale, California 94086, USA, (e-mail: \{hualiu,pxli\}@fortinet.com)}

\begin{abstract}

We investigate deep learning based omni intrusion detection system (IDS) for supervisory control and data acquisition (SCADA) networks that are capable of detecting both temporally uncorrelated and correlated attacks. Regarding the IDSs developed in this paper, a feedforward neural network (FNN) can detect temporally uncorrelated attacks at an {F$_{1}$} of {99.967${\pm}$0.005\%} but correlated attacks as low as {58${\pm}$2\%}. In contrast, long-short term memory (LSTM) detects correlated attacks at {99.56${\pm}$0.01\%} while uncorrelated attacks at {99.3${\pm}$0.1\%}. Combining LSTM and FNN through an ensemble approach further improves the IDS performance with {F$_{1}$} of {99.68${\pm}$0.04\%} regardless the temporal correlations among the data packets.  
\end{abstract}


\section{Introduction}
\IEEEPARstart{S}{upervisory} control and data acquisition (SCADA) is a well established industrial system to automate/monitor processes and to gather data from remote or local equipment such as programmable logic controller (PLC), remote terminal units (RTU) and human-machine-interfaces (HMI), etc. SCADA became popular in the 60's for power plants, water treatment~\cite{1}, and oil pipelines~\cite{2}, etc., which were usually disconnected from the Internet and made use of hardware devices running proprietary protocols. The network was secured from harmful attacks because of its obscurity, thus security means were barely implemented. However, as more and more SCADA systems are adopting its Modbus protocol over TCP and are accessible via the Internet, they are vulnerable to cyberattacks. In 2010, Stuxnet~\cite{stuxnet} was spread over the world and damaged Iranian nuclear power plants. Since then, the need for industrial network security became urgent.

To safeguard SCADA networks, an intrusion detection system (IDS) needs to be implemented. IDS can be signature-based or anomaly-based. Traditionally, signature-based IDS is the mainstream to detect SCADA attacks. It identifies specific patterns from traffic data to detect the malicious activities and can be implemented as policy rules in IDS software such as Snort~\cite{snort,snort_c}. Ref.~\cite{4} investigates a set of attacks against Modbus and designs rules to detect attacks. Ref.~\cite{SRID} proposes a state-relation-based IDS (SRID) to increase the accuracy and decrease the false negative rate in denial-of-service (DoS) detection. However, these detection methods are too complicated and only valid for specific scenarios. Overall, as discovered in the previous research, signature based IDS is only efficient at finding known attacks and its performance relies heavily on the experts' knowledge and experiences. 

An anomaly-based IDS~\cite{AIDS} overcomes these challenges by introducing machine learning to identify attack patterns from data. It is also widely used in other applications such as mobile data misuse detection~\cite{aids_mobile}, software~\cite{aids_c} and wireless sensor security~\cite{aids_w}. Several machine learning algorithms are proposed to develop anomaly-based IDS. Linda \textit{et al.}~\cite{linda2009neural} tailored a neural network model with error-back propagation and Levenberg-Marquardt learning rules in their IDS. Rrushi and Kang~\cite{rrushi2009detecting} combined logistic regression and maximum likelihood estimation to detect anomalies in process control networks. Poojitha \textit{et al.}~\cite{poojitha2010intrusion} trained a feedforward neural network (FNN) to classify intrusions on the KDD99 dataset and the industrial control system dataset. Zhang \textit{et al.}~\cite{zhang2011distributed} used support vector machine and artificial immune system to identify malicious network traffic in the smart grid. Maglaras and Jiang~\cite{maglaras2014intrusion} developed a one-class support vector machine module to train network traces off-line and detect intrusions on-line. All these machine learning algorithms are excellent in observing the pattern of attacks from the in-packet features. None of them, however, takes into account of the temporal features between packets and thus will not perform well on attacks such as DoS which has strong temporal dependence.

DoS attacks are among the most popular attacks to slow down or even crush the SCADA networks. Most of the devices in SCADA operate in low power mode with limited capacity and are vulnerable to DoS~\cite{8}.  Up to date, various DoS types, including spoofing~\cite{spoof}, flooding and smurfing~\cite{smurf}, etc., have been reported. Among all types of DoS, flooding DoS is widely-exploited where hackers send a massive number of packets to jam the target network. In~\cite{13}, the author exploits TCP syn flooding attack against the vulnerability of TCP transmission using hping DoS attack tool. Flooding DoS, along with all other DoS, is difficult to detect because the in-packet features extracted from each data packet may not display any suspicious pattern~\cite{10}. 

Similar to DoS, man-in-the-middle (MITM) is another attack that is hard to detect from observing the in-packet features. It will be more efficient to detect them by observing the inter-packet patterns in time domain.

Anomaly-based IDS on DoS and MITM becomes popular along with the advances of machine learning. For example, in~\cite{AAKR}, an auto-associative kernel regression (AAKR) coupled with the statistical probability ratio test (SPRT) is implemented to detect DoS. The result is not satisfactory because the regression model does not take the temporal signatures of DoS into consideration. In~\cite{3}, FNN is used to classify abnormal packets in SCADA with 85\% accuray for MITM-based random response injection and 90\% accuracy for DoS-based random response injection attacks but 12\% at Replay-based attacks. The author exploits various attacks including DoS attacks and man-in-the-middle (MITM) attacks in the testbed built in Modbus/RTU instead of Modbus/TCP. In~\cite{OCSVM}, the authors propose one-class support vector machine (OCSVM) combined with k-means clustering method to detect the DoS. They set flags on every 10 packets to reflect the relationships of time series. But the handcrafted features may be easily by-passed by expert attackers.

To detect temporally correlated attacks such as flooding DoS and MITM, one should capture the temporal anomaly from these attacks. However, those above mentioned IDS are not designed to extract temporal patterns from packets sequence. A more practical approach is to implement an IDS with the capacity of time series analysis.

Recurrent neural networks (RNN) are the machine learning models that incorporate the recognition of temporal patterns. Among all RNN models, long short-term memory (LSTM) gains its popularity from speech recognition~\cite{Speech}, music composition~\cite{music} and to machine translation~\cite{translation}. It is designed to predict future events according to the information in the previous time steps and suitable for detecting attacks with temporal correlation. For example, Ref.~\cite{LSTM_ddos} applied LSTM for distributed DoS with high success rate. In~\cite{6} the authors also developed a time-series anomaly detector based on LSTM~\cite{Hochreiter:1997:LSM:1246443.1246450} networks to enhance the performance of IDS and apply this framework to the dataset in~\cite{5}. But the number of DoS attacks in the dataset is relatively small and the time interval for the DoS attack in this dataset is too long, making the detection inefficient.

Despite of the excellent performance in detecting temporally correlated attacks such as DoS and MITM, the capacity of RNN to detect temporally uncorrelated attacks is limited compared to other types of machine learning algorithms such as FNN. In this paper, utilizing the advantages of both RNN and FNN while avoiding their disadvantages, we implement an omni IDS that can detect all attacks regardless of their temporal dependence. On a SCADA testbed~\cite{8}, we demonstrate that our IDS reaches the highest performance against all attacks compared to those that employ RNN or FNN alone.

\section{SCADA Testbed and Data Synthesize} \label{sec:simulatednetwork}

Our IDS is tested on a simulated SCADA testbed. A simulated network has the advantage of being easy to maintain, change and operate and is less costly than a real device network. A software testbed, which simulates a SCADA industry network and emulates the attacks was built by L. Zhang~\cite{8} on the basis work of T. Morris~\cite{MORRIS201188}. In the past, several preliminary researches on SCADA security had been conducted on this testbed~\cite{Wang_Hierarchical,shivamThesis}. The attack target is a simple SCADA network, consisting of two tanks using Modbus over TCP. The liquid level of the tanks is controlled by pumps and measured by sensors via Modbus control information. The purpose of this network is to attract hackers and study possible defense methods. Such a system is called Honeypot, as it fools the attacker while exploiting his behaviour. This tank system is developed by the MBLogic HMIBuilder and HMIServer toolkit~\cite{mblogic} and has been extended by L. Zhang in~\cite{8}. The HMI's purpose is to pull data from the sensor or send the desired pump speed to the motor periodically.  The back end of the HMI is a PLC while the front end is a web browser.

\begin{figure}[!ht]
\centering
\includegraphics[width=3.5in]{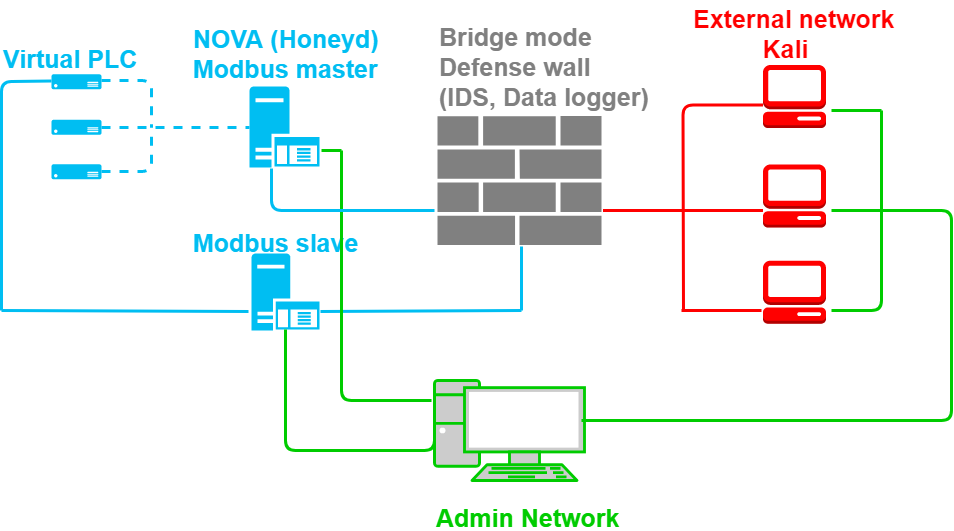}
\caption{Testbed architecture~\cite{8}}
\label{fig:network}
\end{figure}

As this system is simulated, we make use of four virtual machines as shown in Fig.~\ref{fig:network}. The SCADA system runs on a Modbus master and several slaves. On a virtual host called Nova the HMI is deployed, thus we refer to this host as Modbus master. In order to extend the network, some Modbus slaves such as PLCs are simulated by the HoneyD software~\cite{honeyd}. This will provide a more realistic Honeypot. The role of a Modbus slave is to process commands from the master by pulling sensory data about the tank system from the PLCs and sending it back to the master.

The data needed to feed the neural network is generated by an attack machine using a virtual host named Kali. Kali is a Debian-derived Linux host used for penetration testing and features many attack and defense tools. Additional to the message exchange between the Modbus master (Nova) and its slaves we can launch normal traffic mixed with various attacks from Kali. A command line tool, Modpoll~\cite{modpoll}, is used to send Modbus instructions to the PLC which controls sensible tank system variables. An example Modpoll instruction which sends a pump speed of 5 to the system looks like this:

\lstset{language=sh}
\lstset{tabsize=1}

\begin{lstlisting}
	$ modpoll -0 -r 32210 10.0.0.5 5
\end{lstlisting}

The command addresses a simulated PLC with an IP address of 10.0.0.5 and a register address which contains either a threshold value (register 42212 - 42215), the current pump speed (32210) or the tank level (42210,42211), measured by the sensors. Modpoll will send Modbus requests with function code 16 to attempt a write action to the specified registers. By modifying the pump speed the attackers can exceed the allowed tank level and create serious damage to a system. A script on Kali will randomly choose between these normal or malicious Modbus instructions and will launch a Modpoll instruction with another randomly chosen parameter. This will ensure desired distribution of attack/non-attack data.

The traffic will be recorded by the fourth virtual machine referred to as ``Defense Wall'', which operates in the bridge mode and thus is invisible to the attacker. With PyShark we capture the traffic between Nova and Modbus slaves and between the attacker machine Kali and the PLCs. During this process we can label each packet as malicious or normal. 

\subsection{Features extracted from the data packets}\label{appendix_features}
In our testbed, we use a self-developed IDS installed on ``Defense Wall'' to extract 19 features from each data packet captured. They are listed below: 
\begin{enumerate}
\item Source IP address;
\item Destination IP address;
\item Source port number;
\item Destination port number;
\item TCP sequence number;
\item Transaction identifier set by the client to uniquely identify each request;
\item Function code identify the Modbus function used;
\item Reference number of the specified register;
\item Modbus register data;
\item Modbus exception code;
\item Time stamp;
\item Relative time;
\item Highest threshold;
\item Lowest threshold;
\item High threshold;
\item Low threshold;
\item Pump speed;
\item Tank 1 water level;
\item Tank 2 water level.
\end{enumerate}

Here, the ``Relative time'' represents the time in seconds for packets relative to the first packet in the same TCP session. To reduce the periodicity of this feature, we reset it to zero when ``Relative time'' reaches 3,000 seconds. 

In our IDS, we adopt feature scaling of each feature $x$ in the dataset according to
\begin{equation}
x^\prime=\frac{x-\bar{x}}{\sigma_x}
\end{equation}
where $\bar{x}$ and $\sigma_x$ are the mean and standard deviation of original feature $x$ and $x^\prime$ is the re-scaled feature from $x$ with zero mean and unity variance. 

\subsection{Types of attacks in our datasets}\label{appendix_attacks}
\begin{figure}[!t]
\centering
\includegraphics[width=\columnwidth]{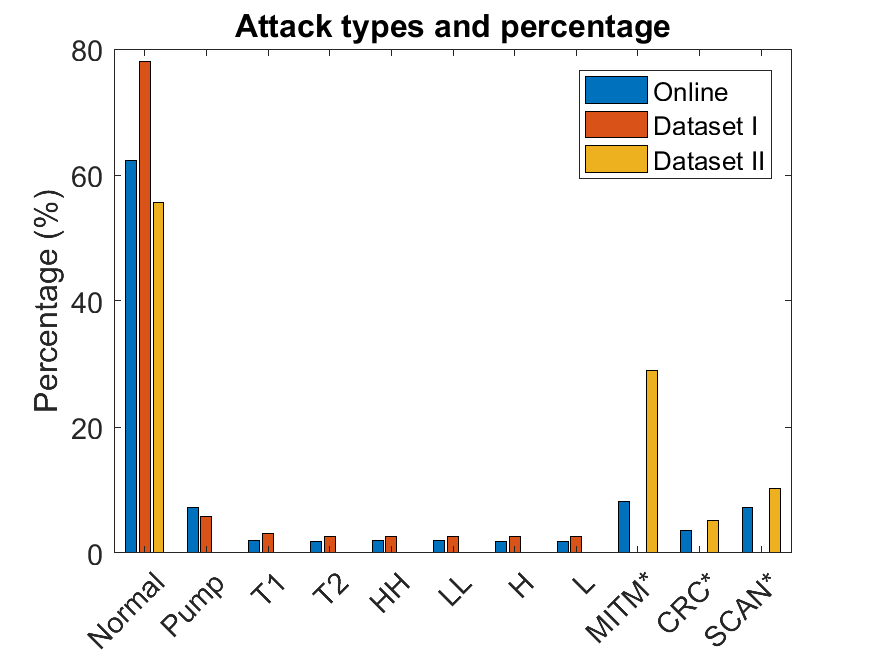}
\caption{Data packet types distribution in Dataset I, II and online script. The ones with a superscript ``*'' are temporally correlated attacks. }
\label{fig:attacks_dist}
\end{figure}
Using our scripts, we created two datasets. As illustrated in Fig.~\ref{fig:attacks_dist}, in addition to ``Normal'' data packets, Dataset I contains attacks that are uncorrelated in time domain while Dataset II contains temporally dependent attacks. Here we have incorporated 10 attacks in our testbed. 7 of them are temporally uncorrelated while the remaining 3 are correlated. The temporally uncorrelated attacks include ``Pump Speed'' (Pump), ``Tank 1 Level'' (T1), ``Tank 2 Level'' (T2), ``Threshold Highest'' (HH), ``Threshold Lowest'' (LL), ``Threshold High'' (H) and ``Threshold Low'' (L) whose detailed descriptions can be found in~\cite{8,MORRIS201188}. 

Among all temporally correlated attacks, two types of flooding DoS attacks are included~\cite{5}. The first labelled as ``Scan flooding'' (SCAN) is to send massive scan command, resulting in increasing latency of communications between the HMI and the sensors in SCADA. The second type labelled as ``Incorrect CRC'' (CRC) is sending massive packets with incorrect cyclic redundancy check (CRC) to cause latency of master. 

Another temporally correlated attack included in this testbed is ``Man-in-the-middle'' (MITM) attack. It is an eavesdropping where the attacker monitors the communication traffics between two parties secretly. Here, the MITM attack is launched by Ettercap~\cite{ettercap} using ARP spoofing~\cite{arp_spoof}. One effective way to detect ARP spoofing is identifying the Media Access Control (MAC) address in layer 2 of OSI model. However, most of Network IDSs (NIDS) do not support the protocols in layer 2 such as ARP and MAC protocols. Even Snort requires an ARP spoof preprocessor~\cite{snort_pre} to collect the MAC address information to detect ARP spoofing. Besides, the victim host of ARP spoofing attack would experience packets retransmissions. For SCADA networks, packet retrasmissions or delay may cause great damages. Therefore, the IDS should raise alert when it detects either MITM attack or packets retransmissions. To make the IDS robust in detecting both MITM and packets retransmissions we remove the MAC address feature which was used for labeling MITM attack from the datasets for training neural networks.

At the first stage, FNN and LSTM IDSs will be trained as binary classifiers that only predict attacks from normal traffic and tested on these datasets separately for performance comparisons. In on-line phases, these two IDSs along with our FNN-LSTM ensemble IDS will be trained as multi-class classifiers by the combined datasets to predict various types of attacks from normal traffics and implemented on the testbed. In addition, we also implement a script that can launch realtime attacks for online testing. The online script will randomly launch normal traffic, temporally uncorrelated and correlated attacks with ratios shown in the table to examine the omni-detection capability of different IDSs.

\section{IDS Implementation}
In this paper, we implemented three IDSs: a conventional FNN, a LSTM and a FNN-LSTM ensemble IDS. Here, we use Keras~\cite{keras} to implement tensorflow~\cite{tensorflow} based machine learning models with AdamOptimizer~\cite{kingma2014adam} to train our model. The structure of these IDSs are detailed in the following subsections.

\subsection{FNN IDS}

\begin{figure}[!t]
\centering
\subfloat[\label{fig:basic1}]{\includegraphics[width=\columnwidth]{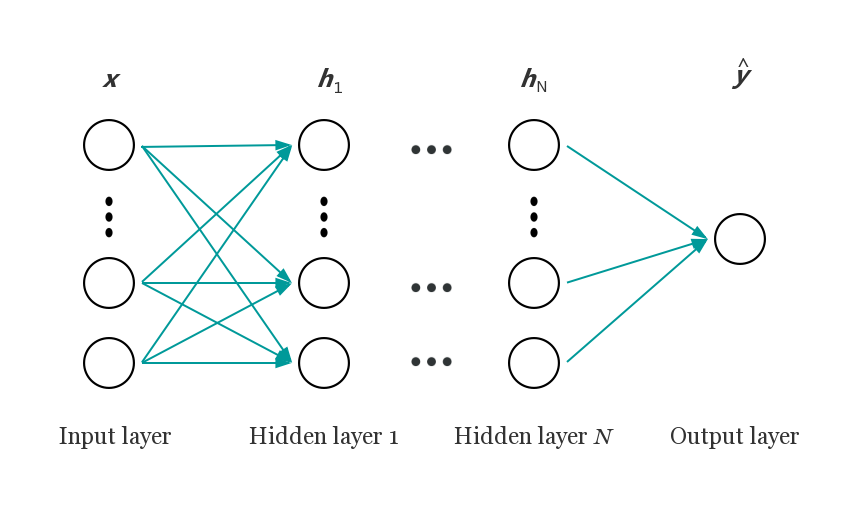}}\\
\subfloat[\label{fig:basic2}]{\includegraphics[width=\columnwidth]{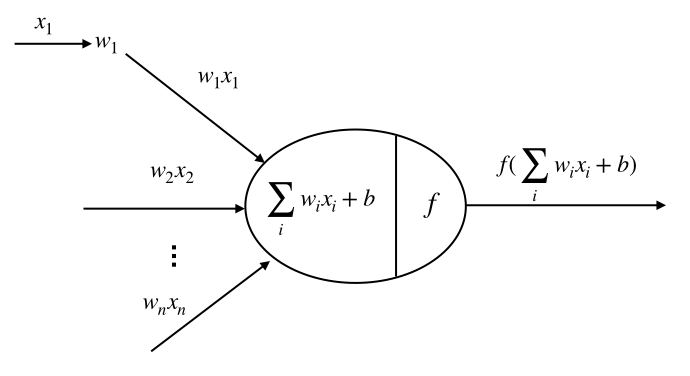}}\\
\caption{(a)The schematics of the FNN IDS (b) Details of each neuron in FNN}
\label{fig:basic}
\end{figure}

The basic structure of the FNN IDS is illustrated in Fig.~\ref{fig:basic}. A typical FNN is formed by an input layer, an output layer and one or more hidden layers in-between. Each layer has a number of neurons that use the neuron outputs from the previous layer as input and produces output to the neurons in next layer. In our case, inputs are the scaled and normalized features extracted from the data packets, and outputs are the predictions of attacks and normal events. Mathematically, the FNN can be expressed as:
\begin{equation}
    \begin{array}{rcl}
\textbf{z}^{(1)}&=&\textbf{W}^{(1)}\textbf{x}+\textbf{b}^{(1)}, \textbf{h}_1=f_h(\textbf{z}^{(1)})\\
\textbf{z}^{(2)}&=&\textbf{W}^{(2)}\textbf{h}_1+\textbf{b}^{(2)}, \textbf{h}_2=f_h(\textbf{z}^{(2)})\\
&...&\\
\textbf{z}^{(N+1)}&=&\textbf{W}^{(N+1)}\textbf{h}_N+\textbf{b}^{(N+1)}, \hat{\textbf{y}}=\textbf{z}^{(N+1)}
\end{array}
\end{equation}
where $N$ is the number of hidden layers, $f_h$ is the ReLU activation function, and $\textbf{W}^{(1)},\textbf{W}^{(2)},..., \textbf{W}^{(N+1)}$, $\textbf{b}^{(1)},\textbf{b}^{(2)}, ..., \textbf{b}^{(N+1)}$ are the parameters to be trained. Here we use softmax cross entropy as our loss function, which can be expressed as  
\begin{equation}
    f_L(\hat{\textbf{y}},\textbf{y})=-\sum_{i=1}^{C}\textbf{y}_{i}\log(f_{s}(\hat{\textbf{y}_i}))
\end{equation}
where   $\hat{\textbf{y}}$ is the predicted label and $\textbf{y}$ the ground truth. $C$ is the number of all possible classes, $\textbf{y}_{i}$ and $\hat{\textbf{y}_i}$ are the actual and predicted labels that belongs to class $i$, and $f_{s}$ is the softmax function.

\subsection{LSTM IDS}
\begin{figure}[!t]
\centering
 \subfloat[\label{fig_sim}]{
\includegraphics[width=3.5in]{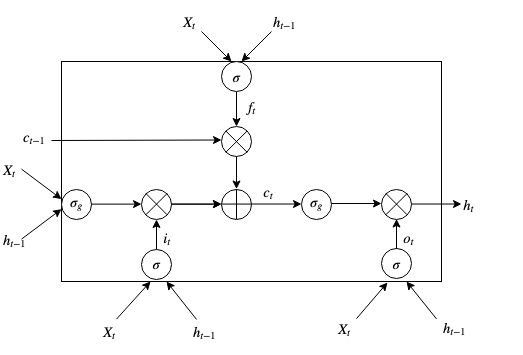}}
\\
\subfloat[\label{LSTM}]{\includegraphics[scale = 0.5]{LSTM_network.png}}
\caption{The structure of (a) single LSTM cell, (b) LSTM Network.}
\end{figure}

 The LSTM is built on a collection of single LSTM cells~\cite{5}. The structure of single LSTM cells is as Fig.~\ref{fig_sim}. Each LSTM cell has 3 gates: input gate, forget gate and output gate. The input gate selects useful information and push it to the cell. The irrelevant information will be discarded in forget gate. The output gate outputs the activation state \(o_t\). A hidden state vector \(h_t\) is transferred to the next time steps. 

The following equations represent the processes of a single LSTM cell:
\begin{equation}
    \begin{array}{rcl}
\textbf{f}_t &=& \sigma(\textbf{W}_fx_t+\textbf{U}_fh_{t-1}+\textbf{b}_f)\\
\textbf{i}_t &=& \sigma(\textbf{W}_ix_t+\textbf{U}_ih_{t-1}+\textbf{b}_i)\\
\textbf{o}_t &=& \sigma(\textbf{W}_ox_t+\textbf{U}_oh_{t-1}+\textbf{b}_o)\\
\textbf{c}_t &=& \textbf{f}_t\circ \textbf{c}_{t-1}+\textbf{i}_t\circ \sigma_g(\textbf{W}_cx_t+\textbf{U}_ch_{t-1}+\textbf{b}_c)\\
\textbf{h}_t&=&\textbf{o}_t\circ \sigma_g(\textbf{c}_t)
\end{array}
\end{equation}
where $\sigma_g$ is hyperbolic tangent function and $\sigma$ is sigmoid function. $\circ$ is the element-wise product notation. $W$, $U$, $b$ are the weight matrix for the gates.

Shown in Fig.~\ref{LSTM}, the LSTM IDS includes two LSTM layers with 10 LSTM cells in each layer. An activation layer with sigmoid activation function is placed after the last LSTM layer.  The ${\{x_1,x_2,...,x_t}\}$ vector is the input vector containing features of packets within $t$ time steps. The dataset is reshaped in this format and fit into the LSTM model. In our model, we set $t=10$. The loss function in this model is binary cross entropy and the optimizer is Adam optimizer~\cite{Adam}.

\subsection{FNN-LSTM Ensemble IDS}\label{sub_ensemble}
\begin{figure}[!t]
\centering
\includegraphics[scale=0.4]{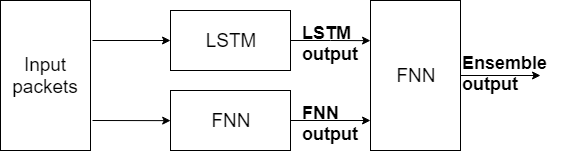}
\caption{Ensemble Model.}\label{fig:LC_DoS}
\end{figure}
Our FNN-LSTM ensemble IDS aims to combine the advantages of both FNN and LSTM while avoiding their weaknesses~\cite{ensemble}. The schematics of this model is as shown in Fig.~\ref{fig:LC_DoS}. In this model, the data packet features are fed into FNN and LSTM simultaneously to predict attacks as a multi-class classifier. The output labels of both are concatenated as the input of a multilayer perceptron, which through training, is capable of voting for the best prediction of the data packet under investigation.

\section{Experiment and Result}

To demonstrate their capability for detecting attacks with/without temporal correlation, we first implement FNN and LSTM IDSs to establish references for comparison. At this stage, the IDSs only conduct binary classification to predict if the data packet under investigation is normal (labeled as ``0'') or attack (labeled as ``1''). Consequently, sigmoid function
\begin{equation}
    \sigma(z) = \frac{e^z}{1+e^z}
\end{equation}
is selected as the activation function. Here, $z$ is the output of the previous LSTM layer. 

\subsection{Hyper parameters tuning}
Both IDSs are trained using 70\% of the randomly chosen samples from the two datasets and tested with the remaining 30\% samples following the 10-fold training/testing procedure so that the average and standard deviation of figures of merits including precision, recall and $\mathrm{F_1}$ can be used for evaluation. 

To determine the number of hidden layers necessary for our FNN, we computed $\mathrm{F_1}$ with 0, 1 and 2 hidden layers where the values of 99.22\%, 99.96\% and 99.97\% are obtained respectively. As shown, employing 1 hidden layers in FNN will increase the $\mathrm{F_1}$ by more than 7\% while using 2 hidden layers the improvement is minimal. Therefore, we select 1 hidden layer in our FNN implementation.

In addition, to circumvent overfitting, we further adopted early stop procedure in FNN such that the optimization stops when the number of epochs whose relative differences of loss between consecutive ones are less than $\mathrm{10^{-6}}$ reaches 35~\cite{es}. Similarly, LSTM adopts early stop if either maximum epochs reach 3.

In implementation of LSTM, we connect 10 LSTM cells in input layer where the features from 10 consecutive data packets are entered into the cells to predict if the last packet is normal or an attack. In training, we adopt mini-batch with a batch size of $\mathrm{1,000}$.

\subsection{Detection of temporally uncorrelated attacks}

\begin{figure}
\centering
\includegraphics[scale=0.65]{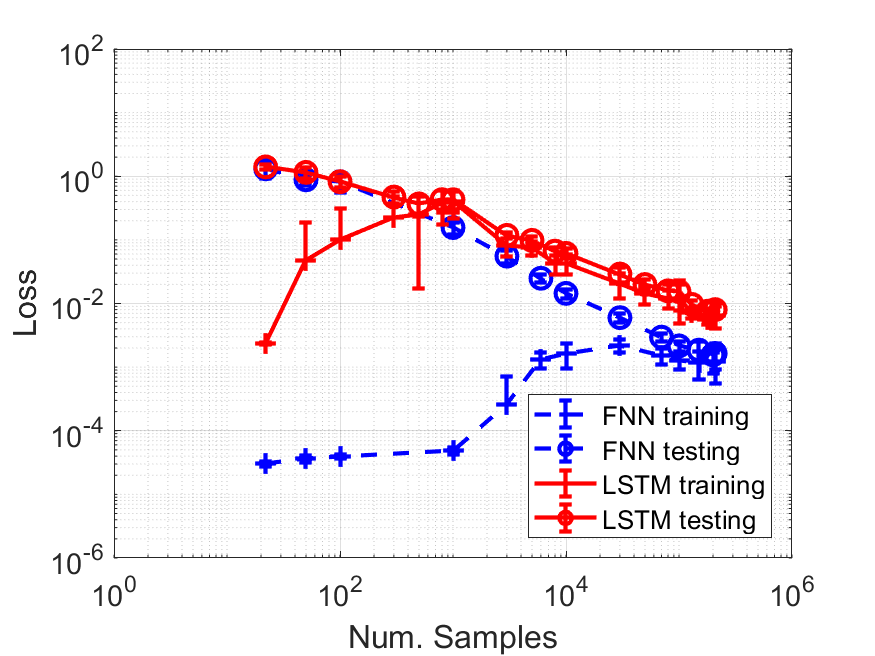}
\caption{Learning Curves of FNN and LSTM using temporally-uncorrelated-attacks dataset (Dataset I).}\label{learning_dsetI}
\end{figure}

We exploit the Dataset I described in Section~\ref{sec:simulatednetwork} to compare the detection capability of FNN and LSTM for temporally uncorrelated attacks.  To verify the models, learning curves are plotted in Fig.~\ref{learning_dsetI} where training and testing losses as a function of training samples are plotted. Here the average value and standard deviation after 10 fold training/testing are represented by circle markers and error bars respectively. As shown, with training samples exceeding 40,000, FNN training and testing losses (blue dashed lines) start to converge while LSTM (red solid lines) converges at sample size larger than 60,000. Overall, it confirms that the number of samples in Dataset I is sufficient for the training and testing of our IDS.

\begin{table}[!t]
\normalsize
\centering
\caption{\label{tab:test_nonDoS}Comparison of the temporally-uncorrelated-attacks detection.}
\begin{tabular}{|c|c|c|c|}
\hline
 & { Precision} & { Recall} & { F}$_{1}$ \\\hline
{ FNN} & $\mathrm{99.996{\pm}0.006}$ & $\mathrm{99.84{\pm}0.05}$ & $\mathrm{99.92{\pm}0.03}$ \\\hline
{ LSTM} & $\mathrm{99.88{\pm}0.06}$ & $\mathrm{98.7{\pm}0.4}$ & $\mathrm{99.3{\pm}0.1}$  \\\hline

\end{tabular}
\end{table}

\begin{table}[!t]
\normalsize
\caption{Confusion matrices of temporally-uncorrelated-attacks detection using Dataset I (averaged over 10 trials)}\label{tab:confusion_nonDoS}
\centering
\begin{tabular}{|c|c|c|c|c|}
\cline{4-5}
\multicolumn{3}{c|}{}&\multicolumn{2}{c|}{{\bf Predicted}}  \\
\cline{4-5}
\multicolumn{3}{c|}{}&{\bf Normal} & {\bf Attacks} \\
\hline
\multirow{6}{*}{\vspace{0.3in}\rotatebox[origin=c]{90}{{\bf Actual}}} &  \multirow{3}{*}{\vspace{0.1in}{\bf Normal}} & {\bf FNN} & $\mathrm{69,845.4}$ & $\mathrm{0.6}$ \\\cline{3-5}
&&{\bf LSTM} & $\mathrm{69,902.2}$& $\mathrm{22.8}$ \\\cline{2-5}
&\multirow{3}{*}{\vspace{0.1in}{\bf Attacks}} & {\bf FNN} & $\mathrm{30.7}$ & $\mathrm{19,741.3}$ \\\cline{3-5}
&&{\bf LSTM}& $\mathrm{241.9}$& $\mathrm{19,448.1 }$ \\\hline
\end{tabular}
\end{table}

After the IDSs are trained, we use 30\% of samples in Dataset I for 10 fold testing. Also shown in Table~\ref{tab:confusion_nonDoS} and~\ref{tab:test_nonDoS}, on average, for FNN, only 0.6 of the 69,846 normal datapackets are mislabelled as attacks while only 30.7 out of 19,771 actual attacks are mislabelled as normal traffic, yielding the precision, recall and $\mathrm{F_1}$ to be $\mathrm{99.996{\pm}0.006\%}$, $\mathrm{99.84{\pm}0.05\%}$, and $\mathrm{99.92{\pm}0.03\%}$. In comparison, LSTM mislabelled 22.8 normal packets as attacks and 241.9 attacks as normal packets, resulting the figures of merits to be $\mathrm{99.88{\pm}0.06\%}$, $\mathrm{98.7{\pm}0.4\%}$ and $\mathrm{99.3{\pm}0.1\%}$.  The comparison demonstrates that FNN outperformed LSTM in detecting temporally uncorrelated attacks where recognition of the in-packet feature patterns is critical. 

\subsection{Detection of temporally correlated attacks}

\begin{figure}[!t]
\centering
\includegraphics[scale=0.65]{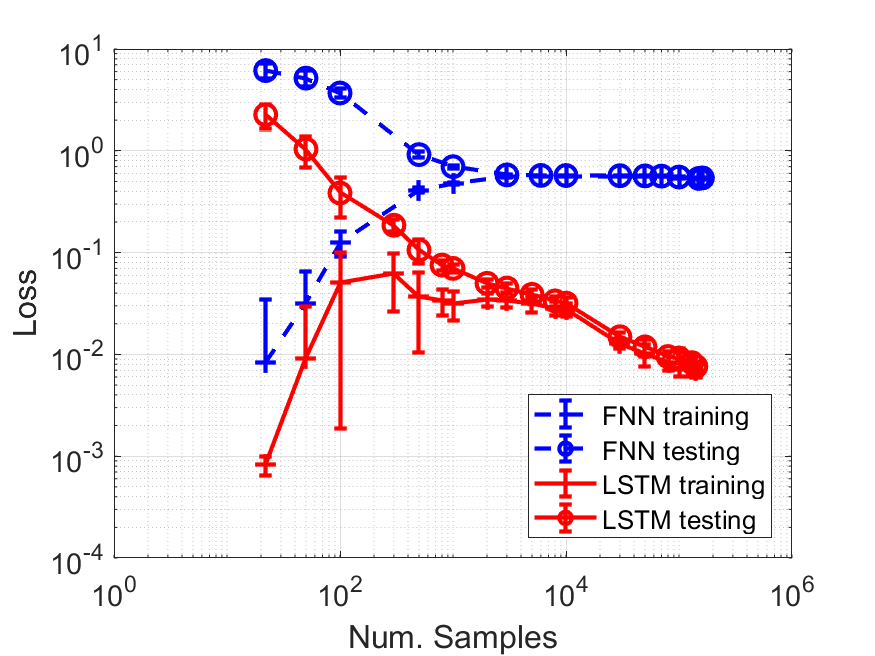}
\caption{Learning Curves of FNN and LSTM using temporally-correlated-attacks dataset (Dataset II).}\label{fig:LC_DoS}
\end{figure}

\begin{table}[!t]
\normalsize
\caption{Comparison of temporally correlated attacks $\mathrm{(\%)}$}\label{tab:test_DoS}
\centering
\begin{tabular}{|c|c|c|c|}
\cline{2-4}
\multicolumn{1}{c|}{} & {\bf Precision} & {\bf Recall} & {\bf F}$_{1}$ \\\hline
{\bf FNN}  & $\mathrm{73{\pm}2}$ & $\mathrm{49{\pm}4}$ & $\mathrm{58{\pm}2}$ \\\hline
{\bf LSTM}  & $\mathrm{99.60{\pm}0.01}$ & $\mathrm{99.52{\pm}0.02}$ & $\mathrm{99.56{\pm}0.01}$\\\hline
\end{tabular}
\end{table}

\begin{table}[!t]
\normalsize
\caption{Confusion matrix of temporally correlated attacks}\label{tab:confusion_DoS}
\centering
\begin{tabular}{|c|c|c|c|c|}
\cline{4-5}
\multicolumn{3}{c|}{}&\multicolumn{2}{c|}{{\bf Predicted}}  \\
\cline{4-5}
\multicolumn{3}{c|}{}&{\bf Normal} & {\bf Attacks} \\
\hline
\multirow{6}{*}{\vspace{0.3in}\rotatebox[origin=c]{90}{{\bf Actual}}} &  \multirow{3}{*}{\vspace{0.1in}{\bf Normal}} & {\bf FNN} & $\mathrm{28,668.3}$ & $\mathrm{5,044.7}$\\\cline{3-5}
&&{\bf LSTM}& $\mathrm{33,504.0}$ & $\mathrm{105.0}$ \\\cline{2-5}
&\multirow{3}{*}{\vspace{0.1in}{\bf Attacks}} & {\bf FNN} & $\mathrm{13,510.4}$ & $\mathrm{13,169.6}$\\\cline{3-5}
&&{\bf LSTM}& $\mathrm{128.4}$ & $\mathrm{26,652.6} $ \\\hline
\end{tabular}
\end{table}

%\IEEEpubidadjcol
 In this subsection FNN and LSTM are re-trained and tested using Dataset II for comparison of their temporally correlated attacks detection comparison. Again the learning curves in Fig.~\ref{fig:LC_DoS} shows that both FNN (blue dashed lines) and LSTM (red solid lines) converge at training samples exceeding 10,000 while LSTM clearly shows lower testing loss. This confirms  the sufficiency of our dataset to generalize the IDS models.

The performance of each model is compared in Table~\ref{tab:test_DoS} and \ref{tab:confusion_DoS}. As shown, FNN is inefficient in detecting temporally correlated attacks with precision, recall and $\mathrm{F_1}$ scores as low as $\mathrm{73{\pm}2\%}$, $\mathrm{49{\pm}4\%}$ and $\mathrm{58{\pm}2}$ respectively. In particular, 5,044.7 out of 33,713 normal packets are mislabelled to attacks while 13,510.4 out of 26,680 actual attacks are mislabelled to normal traffic. It is evident that the poor performance of FNN is caused by its inability to inter-packet features. In contrast,  LSTM displays an outstanding performance on the corresponding figures of merits to be  $\mathrm{99.60{\pm}0.01\%}$,  $\mathrm{99.52{\pm}0.02\%}$ and  $\mathrm{99.56{\pm}0.01\%}$ where only 105.0 normal packets are mislabelled as attacks and 128.4 attacks packets are mislabelled as normal traffic. As expected, LSTM outperforms FNN in detecting temporally correlated attacks due to its inherent nature to observe data pattern in time domain.

\subsection {Omni attacks detection}

\begin{table}[!t]
\normalsize
\caption{Macro-average comparison of omni-attacks detection}\label{omni_macro}
\centering
\begin{tabular}{|c|c|c|c|}
\cline{2-4}
\multicolumn{1}{c|}{} &{\bf Precision}	& {\bf Recall}	& {\bf F}$_{1}$ \\
\hline
{\bf FNN}&$\mathrm{88{\pm}1}$			&$\mathrm{89.2{\pm}0.8}$&$\mathrm{87.4{\pm}0.6}$       \\
\hline
{\bf LSTM} &$\mathrm{99.54{\pm}0.03}$&$\mathrm{99.01{\pm}0.07}$&$\mathrm{99.27{\pm}0.05}$\\\hline
{\bf Ensemble} &$\mathrm{99.76{\pm}0.05}$&$\mathrm{99.57{\pm}0.03}$&$\mathrm{99.68{\pm}0.04}$\\\hline
\end{tabular}
\end{table}

\begin{figure}
\centering
\subfloat[\label{omni_precision}]{
\includegraphics[scale=0.6]{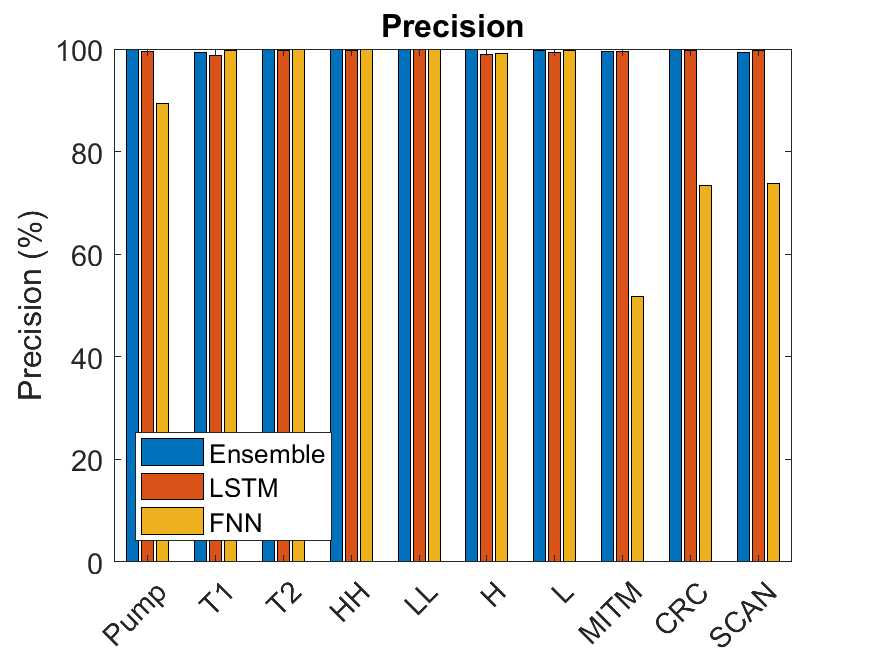}
}\\
\subfloat[\label{omni_recall}]{
\includegraphics[scale=0.6]{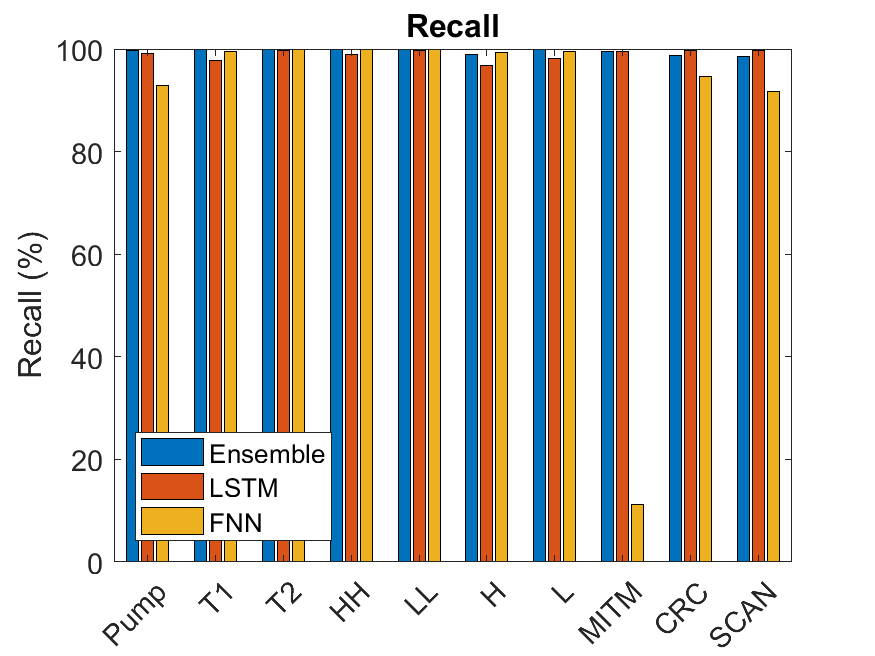}
}\\
\subfloat[\label{omni_F1}]{
\includegraphics[scale=0.6]{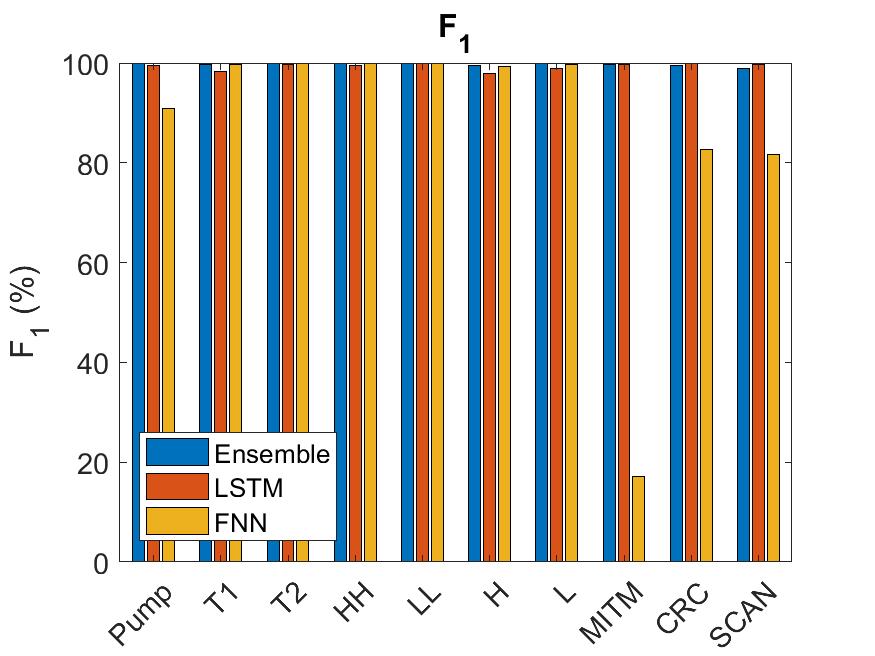}
}
\caption{(a) Precision, (b) Recall and (c) $\mathrm{F_1}$ of individual attacks in omni-attacks detection.}\label{fig:LC_DoS}
\end{figure}

Recognizing the mutual strength of FNN and LSTM IDSs in detecting temporally correlated and uncorrelated attacks, we here combine the advantages of both for an omni attacks detector through ensemble approach. The structure of FNN-LSTM ensemble is described in Subsection~\ref{sub_ensemble}. To implement, we first remodelled FNN and LSTM to multi-class classifiers so that different attacks can be distinguished. Dataset I and II are combined and used to train FNN and LSTM independently. The outputs of both are combined to form the input features of a multilayer perceptron for training. After training, FNN, LSTM and FNN-LSTM ensemble IDSs are integrated into our SCADA testbed to detect and classify attacks.  The traffic is generated online using the script that generates a pre-determined ratio of normal, temporally correlated and uncorrelated attacks as described in Fig~\ref{fig:attacks_dist}. To estimate the figures of merits, we evenly divide the predicted labels to 10 portions and compute the average and standard deviation of macro-averaged precision, recall and $\mathrm{F_1}$. As shown in Table~\ref{omni_macro}, among all the three IDSs, the FNN achieve lowest performance with macro-averaged figures of merits of $\mathrm{88{\pm}1\%}$, $\mathrm{89.2{\pm}0.8\%}$ and $\mathrm{87.4{\pm}0.6\%}$ while LSTM reaches $\mathrm{99.54{\pm}0.03\%}$, $\mathrm{99.01{\pm}0.07\%}$ and $\mathrm{99.27{\pm}0.05\%}$.
In contrast, the FNN-LSTM ensemble IDS further outperforms both with figures of merits to be $\mathrm{99.76{\pm}0.05}$, $\mathrm{99.57{\pm}0.03}$ and $\mathrm{99.68{\pm}0.04}$.  Detailed analysis in Fig.~\ref{fig:LC_DoS} further confirms that the under-performance of FNN (yellow bars) are due to the mislabels of temporally correlated attacks (MITM, CRC and SCAN) while the performance of LSTM (red bars) by temporally uncorrelated attacks (``Pump Speed (Pump)'', ``Tank 1 Level (T1)'', and ``Threshold High (H)'', etc.). Overall, the FNN-LSTM ensemble demonstrates a consistent out-performance over them in all types of attacks.

\section{Conclusion}
In this paper we demonstrated that the FNN-LSTM ensemble IDS can detect all types of cyberattacks regardless of the their temporal relevance. In opposite, FNN only performance well in temporally uncorrelated attacks and LSTM is relatively weak in uncorrelated attacks. In future research we will further improve our model through field trials.

\ifCLASSOPTIONcaptionsoff
  \newpage
\fi

\appendices
\section*{Acknowledgment}
\end{document}